\documentclass{ws-procs9x6}

\newcommand{\be}{\begin{eqnarray}}
\newcommand{\ee}{\end{eqnarray}}
\newcommand{\bib}{\bibitem}

\newcommand{\rar}{\rightarrow}

\newcommand{\mx}{m_x}
\newcommand{\taux}{\tau_x}
\newcommand{\mpl}{m_{Pl}}

\begin{document}

\title{
 Big Bang and Heavy Particles 
}

\author{A.D. DOLGOV
}

\address{
INFN, sezione di Ferrara, Via Paradiso, 12 - 44100 Ferrara, Italy; \\
ITEP, Bol. Cheremushkinskaya 25, Moscow 113259, Russia\\
ICTP, Strada Costiera 11, 31014 Trieste, Italy;\\
E-mail: adolgov@fe.infn.it}

\maketitle

\abstracts{
A possibility of existence of ultra-heavy (quasi)stable particles,
mechanisms leading to their large life-time, their production in 
the early universe, and cosmological manifestations are reviewed. 
}

\section{Introduction \label{s-intr}}

We know from experiment that there are only a few stable particles,
$e$, $p$, $\nu$, and $\gamma$ 
which are relatively light. It is an open question if there exist 
heavier stable or quasi-stable particles. Possibly they have not yet been
discovered because the energies of available accelerators are not large enough
for their production. On the other hand, we have ``poor man accelerator'',
big bang, and believe that in the early universe all kind of particles
with the mass of the order of the temperature of primeval plasma, or even
considerably higher than that, could be produced. If these particles are
stable or long-lived we may hope to see some signatures of them today.
Moreover they can be helpful for solution of some cosmological puzzles.

In what follows I am going to address the following questions:\\
1. Why do we need such heavy (quasi)stable particles, HqSP? (Below we will
use for them either this abbreviation or just denote them as $x$.)\\
2. What type of objects can they be?\\
3. How HqSP could be created in the early universe?\\
4. What mechanism  makes them (quasi)stable?

We do not have rigorous answers to these questions and
in many cases one has to rely on reasonable hypotheses or, as 
is fashionable to say now, conjectures. 

\subsection{Why may we need them? \label{ss-why}}

There are several places where existence of HqSP, may help:\\
If they have mass about $10^{13}$ GeV and life-time of the order of 
or larger than the universe age, $\taux > t_U \approx 10^{10}$ years, 
their decays may solve the mystery of ultrahigh energy cosmic 
rays\cite{bkv} observed beyond the Greisen-Zatsepin-Kuzmin 
cut-off~\cite{gzk}. For recent reviews see e.g.~\cite{uhecr-rev}.\\
A very heavy particle with mass in the range $10^{12}-10^{16}$ GeV and
either absolutely stable or with life-time larger than the universe age 
was suggested in ref.~\cite{wimpz} as a candidate for the
cosmological dark matter.\\
Heavy long-lived particles, though not as long as the universe age, 
could be very efficient in baryogenesis if their decay does not conserve
baryonic charge. The life-time should be much larger than the inverse
Hubble parameter at $T\sim m_x$, i.e. $\taux < m_{Pl}/\mx^2$, to
ensure strong deviation from thermal equilibrium but shorter than 
roughly 1 sec to respect the standard big bang nucleosynthesis.\\
Charged heavy stable particles could catalyze thermonuclear 
reactions~\cite{ads-frank} and make feasible thermonuclear power 
stations if we find a way to produce or dig them out in sufficiently 
large quantities.\\
Even more efficient for power production could be magnetic monopoles
efficiently catalyzing proton decay~\cite{rub-cat}. Finding enough monopoles
could solve energy problems for almost infinitely long time, even after 
the Sun exhausts its nuclear power and dies. 

\subsection{What kind of objects could they be? \label{ss-what}}

There are three (or maybe four) logical possibilities:\\
Normal elementary particles, with the
size of the order of their Compton wave 
length, $l_x\sim 1/\mx$. We can take their mass as a free parameter
and have to understand why such particles are (quasi) stable despite
being heavy. One well known type of heavy (but not very heavy) hypothetical
stable (or long-lived) particle is the lightest supersymmetric particle, 
LSP. They are known to be stable if R-parity is rigorously conserved.
Their mass in minimal supersymmetric extension of the standard model is 
however, not as huge as mentioned above, 
from $\sim 100$ GeV up to a few TeV.
In this mass range LSP is a good candidates for dark matter particles but 
it is not excluded that supersymmetry is realized at much higher energy 
scale and in this case the masses may be much larger.

Magnetic monopoles already mentioned above are not really elementary 
particles because their size $1/(\alpha m)$ is much larger than their
Compton wave length. Still they are pretty small. For GUT monopoles with
$m\approx 10^{17}$ GeV the size is about $10^{-29}$ cm. 

Known non-topological solitons are usually very heavy and large and hence 
we will not consider them as elementary particles. 

Interesting objects are mini black holes (BH) with the mass of the order
of the Planck mass, 
$m_{BH}\sim m_{Pl} = 1.22\cdot 10^{19}{\rm GeV} = 2.18 \cdot 10^{-5}$ g.
In multidimensional models with TeV scale gravity~\cite{tev-grav}
the mass of such black holes could be much smaller, down to TeV.

\section{Stability \label{s-stab}}

\subsection{Mini black holes \label{ss-min-bh}}

Bigger mass black holes evaporate semi-classically through the Hawking process
but nothing is known about quantum mini-BH with mass equal to the
Planck mass. For them unknown effects of quantum gravity are 100\% important.
It is not yet agreed upon if they are stable or not. The answer is probably
determined by the ultraviolet behavior of quantum gravity. If the latter
is UV-free than mini-BH's could quite possibly be stable. 

On the other hand, as the saying goes ``anything which is not forbidden is 
allowed''. It is especially true for quantum field theory and if there is 
no special law to forbid decay the particle must be unstable. We do not know  
any such law for mini-BH but the lack of knowledge does not prove that such 
a law does not exist. If mini black holes 
are stable and if gravity becomes strong at TeV scale then there must 
exist stable particles (black holes) with TeV mass.

\subsection{Magnetic monopoles \label{ss-mm}}

Stability of magnetic monopoles is ensured by topology - it is impossible to
unwind the gauge field lines at least in 3-dimensional space. It corresponds 
to conservation of magnetic charge. Though this conservation looks similar
to conservation of electric charge, it is not the same. If photon is
exactly massless then it is impossible to formulate a consistent theory
with non-conserved charge. However, even with massive but very light
photon, satisfying experimental bounds on photon mass~\cite{pdg}, 
the processes with electric charge non-conservation become exponentially
strongly suppressed, as $\exp[-C(E/m_\gamma)^2]$, where $E$ is the 
characteristic energy of the process and $m_\gamma$ is the photon 
mass, as is argued in ref.\cite{ozv}. Hence, most probably, electron 
would be stable even if electric charge is not conserved. 
It is unclear if the same suppression
is valid for magnetic monopoles, or in the case of massive (but very light) 
photons magnetic charge may be non-conserved and magnetic monopoles could
quickly decay.

\subsection{ Heavy elementary particles \label{ss-heavy-part}}

``Normal'' decay width of an elementary particle is $\Gamma_x \sim g^2 \mx$
where $g$ is a coupling constant characterizing interaction leading to
the decay. Usually $g<1$ but not outrageously small. Anomalously small or 
vanishing decay width in all known cases is explained by a conservation
law related to an unbroken symmetry, as e.g. the
discussed above stability
of electron due to conservation of electric charge which, in turn, is 
related to local $U(1)$-symmetry and vanishing mass of the
gauge boson, photon.

Stability of another known stable particle, proton, is prescribed to
conservation of baryonic charge. The corresponding $U(1)$-symmetry
is most probably global and is not associated with a gauge boson.
An idea of possible long range forces related to baryonic charge was 
put forward in ref.~\cite{ly55} and similar idea about leptonic charge 
forces in ref.~\cite{lbo69} (for a review see ref.~\cite{ad-lrf}). 
It was found in these works, however, that a high precision tests
of the equivalence principle put very stringent bounds on possible 
coupling constant of baryo- and lepto- photons: $\alpha_B < 10^{-47}$
and $\alpha_L < 10^{-48}$. Thus, most probably long range forces
induced by baryonic and/or leptonic charges are absent. It is believed
that global symmetries should be broken by gravity at Planck scale.
As was noticed in ref.~\cite{ybz76} and later analyzed in~\cite{akm}
formation of a virtual Planck mass black hole would induce proton 
instability with life-time of the order of 
\be
\tau_p\sim m^4_{Pl}/m_p^5 \sim 10^{45}\,\, {\rm years}.
\label{tau-p}
\ee
In the case of TeV gravity the life-time of proton would be tiny,
$\tau_p \sim 10^{-12}$ sec. However, if the black hole formation
is suppressed then the same would be true for the proton decay. 
One may expect exponential suppression of the production of classical
objects in elementary particle interactions and, if black hole is
indeed such, then gravitationally induced decay would not be effective. 
If this is true then the accelerator production of TeV mass black holes
should be negligible~\cite{mbv-bh}. Thus we are left with the dilemma:
either proton leaves $10^{-12}$ sec or TeV black-holes cannot be produced
at accelerators~\cite{akm}. 

Since Planck mass black holes have the size $r_g \approx 1/m_{BH}$, i.e.
equal to their Compton wave length, they may be considered as elementary
particles and their production is most probably unsuppressed. If so, 
the mass scale of gravity, $m_{gr}$, can be bounded by the long life-time of
proton\cite{pdg}, $\tau_p > 2.1\cdot 10^{29} $ years. Using equation
(\ref{tau-p}) we obtain $m_{gr} > 2\cdot 10^{15}$ GeV, which is not too
far from the normal Planck scale $10^{19}$ GeV.

Very heavy particles which were introduced for explanation of the 
observed ultra high energy cosmic rays (UHECR) should have mass about 
$10^{13}$ GeV. Their decay, induced by gravity, would have life-time about
1 sec, even with normal huge Planck mass. The same would be true
for ultra heavy dark matter particles. Thus such particles could neither
produce UHECR, nor make cosmological dark matter.  

A possible way to make them stable is to assume that they belong to a
non-singlet representation of a local symmetry group $G$. Either they
are charged with respect to a new $U'(1)$ group and new ``photons'',
$\gamma'$ exist, or there is a new non-Abelian group, similar to QCD,
with a high confinement scale. If this group $G$ is unbroken, then
the lightest charged particle must be completely stable. Instability
could be induced by a minuscule breaking of $G$. 

If such ultra-heavy charged particles are completely stable, they nevertheless
may create energetic cosmic rays if their positronium-like bound
states have been formed in the early universe~\cite{hill,dfk} and their 
annihilation is so slow that the life-time of the corresponding ``-onium''
is larger than the universe age. However,
an explicit realization of this idea seems to meet serious problems.
 
On the other hand, if gravitational decay is exponentially suppressed 
by some yet unknown reason then the life-time of such heavy particles 
could easily be in the required range of $\taux \geq t_U$. There is no
reliable answer to what really takes place and thus any hypothesis is 
allowed.

One more comment may be interesting here. If the scale of strong
gravity is about $m_{gr} \sim 1$ TeV then the particle with mass 
$10^{13}$ GeV and size $l_x=1/\mx$ would be inside their Schwarzschild
horizon, 
$r_g = \mx /m_{gr}^2 = (1/\mx) (\mx/m_{gr})^2 > l_x$ and thus they
form black holes. Such black holes would evaporate by the 
Hawking process and have the life-time 
\be
\taux \sim 0.1 \mx^3/m_{gr}^4 \approx 10^2\,{\rm sec} 
 \left(\frac {\mx}{10^{13}\, {\rm GeV}}\right)^3
\left( \frac{1 \,{\rm TeV}}{m_{gr}}\right)^4
\label{tau-bh}
\ee
This arguments make it very plausible that existence of HqSP particles is 
not compatible with TeV-scale gravity. Of course, if the considered heavy
particles have a conserved gauge charge $Q_g$, then such lightest charged 
black hole would not evaporate and would be absolutely stable. However, even
a small mass of the related gauge boson, $m_g$, would make evaporation 
possible, even if the charge is $Q_g$ is strictly conserved~\cite{dmt}.
In the limit of very small $m_g$ the life-time of 
such charged black holes should be of the order $1/m_g$. An interesting
possibility is a non-minimal gauge non-invariant coupling of the gauge
field to the curvature scalar of the form 
\be
L =  g^{\mu\nu}A^{(g)}_\mu A^{(g)}_\nu R
\label{RAA}
\ee
Since $R\sim 1/t_U^2$ 
the life-time of 
$g$-charged black holes would be always of the order of $1/H\sim t_U$.

\section{Production mechanisms \label{s-prod}}

\subsection{Thermal production \label{ss-therm}}

All particles with masses smaller or of the order of temperature
are abundant in cosmic plasma if thermal equilibrium with respect
to them is established. If temperature is small in comparison with 
the particle mass then their abundance is Boltzmann suppressed. The
ratio of massive to massless number densities at $T<m$ is:
\be
n(m) /n (0) \approx \left(m/T\right)^{3/2}\,e^{-m/T}
\label{bolt-suppr}
\ee
The condition for equilibrium is slow cosmological expansion 
in comparison with the reaction rate, 
\be
\dot n/n \sim \sigma n > \dot a /a\equiv H,
\label{equil}
\ee
where $\sigma$ is the interaction cross-section, $a(t)$ is the cosmological 
scale factor, and $H$ is the Hubble parameter. The latter is expressed 
through the total cosmological energy density:
\be 
H = \frac{8\pi\,\sqrt\rho}{3m_{Pl}} = 
\frac{8\pi^3 g_*}{90}\,\frac{T^2}{m_{Pl}}
\label{H-of-T}
\ee
where $g_*$ is the effective number of relativistic species.

If the universe temperature after inflation was low, $T^{(in)}<\mx$,
then the heavy $x$-particles would not be thermally produced. However,
one can outwit the Nature if would-be heavy particles $x$ were massless
at the production time and became massive later as a result of phase
transition which made them massive~\cite{mass}. In such a case
the plasma might be populated by heavy particles with $m>T$ and with
the number density equal to that of 
equilibrium massless particles, $n\sim T^3$.

\subsection{Topological production \label{ss-top}}

Classical objects, i.e. the objects with size bigger than their 
Compton wavelength, $l>1/m$, probably cannot be efficiently created
in thermal bath. For example, 
magnetic monopoles are produced by the so called
Kibble mechanism\cite{kibble}: in causally disconnected regions the 
gauge field lines can be arbitrary winded and end up in monopole 
configuration, roughly one per one Hubble volume, $H^{-3}$,
or Ginsburg correlation volume, $(\lambda \eta)^{-3}$, where $\eta$ is the
vacuum expectation value of the Higgs field and $\lambda$ is the
coupling constant of the quartic self-interaction, 
$\lambda (|\phi|^2-\eta^2)$.

Topological defects could be formed at the end of inflation at the
preheating stage~\cite{top-infl} because of possible phase transitions
(for a review see ref.~\cite{kuz-tka}).

Production of classical objects in particle collisions have not yet been
rigorously calculated. There is a common agreement that production of a
pair of magnetic monopoles in the reaction $e^+e^- \rar M \bar M$ is
exponentially suppressed. On the other hand, monopole annihilation
into all possible particles, $M \bar M \rar all$, is probably quite
efficient with the cross-section $\sigma \sim C/m_M^2$ with a constant
coefficient $C\geq 1$. This process is probably dominated by a large 
multiplicity reactions with the number of produced particles of the order
of $1/\alpha \sim 100$. One may argue that due to time invariance and
related detailed balance condition (even an approximate one
because of a small breaking of time invariance ) the inverse 
process $(100\,\, particles) \rar M\bar M$ would proceed with the same
probability and hence thermal production of monopoles would be
unsuppressed. On the other hand, it looks quite plausible that in
$M\bar M$-annihilation a very special coherent state of final particles
is formed which practically never occurs in thermal bath. If this is
true then the annihilation of monopoles would be efficient, while the
inverse process would be suppressed by a huge entropy factor.

One may encounter the same or similar problems in formation of 
black holes in accelerators, discussed above, and in electro-weak 
baryogenesis~\cite{krs} which proceeds through formation of
classical field configurations, sphalerons. For more detailed
discussion and literature see~\cite{ad-taup}.

\subsection{Particle production by inflaton field \label{ss-prod-infl}}

In the course of inflation the inflaton field $\Phi(t)$ evolves
very slowly but closer to the end the evolution becomes much faster and
$\Phi(t)$, as any time-dependent field, starts to create particles
with which it is coupled. Typically the produced particles have 
the masses which are smaller than or comparable to the characteristic
frequency of the inflaton variation, $\omega \sim \dot \Phi/\Phi$.
The most favorable situation for particle creation is the 
regime when $\Phi$ oscillates around the minimum of its potential:
\be
\Phi(t) = \Phi_0 (t) \cos \left( m_\Phi t + \delta \right)
\label{phi-of-t}
\ee
The interaction Lagrangian of $\Phi$ with usual  particles is of the
type:
\be
{ L}_{int} = g\Phi \bar \psi \psi + \lambda \Phi^2 \chi^* \chi + ...
\label{l-int}
\ee
where $\psi$ and $\chi$ are respectively quantum fermion and boson fields
and the inflaton field $\Phi(t)$ is considered as a classical field.
The coupling constants $g$ and $\lambda$ are supposed to be small, 
otherwise self-interaction of the inflaton induced by loop corrections
would be unacceptably strong. Hence perturbation theory is applicable
for description of the particle production~\cite{perturb}, 
but only in the case
when the masses of the produced particles are smaller than the oscillation
frequency, $\omega = m_\Phi$. In this case the probability of fermion
production per unit time, $\dot n_\psi /n_\Phi $,
is simply the decay width of $\Phi$-boson, 
\be
\Gamma_{pt} = g^2 m_\Phi /8\pi.
\label{gam-pt}
\ee
The energy of each produced fermion is equal to $m_\Phi/2$.
The same result would be true for any trilinear coupling 
$\Phi \chi_1 \chi_2$. For the quartic coupling of eq. (\ref{l-int}) the
production probability is
$\dot n_\chi/n_\Phi \sim \lambda^2 \Phi_0^2 /m_\Phi$ and the energy
of each produced boson being equal to $m_\Phi$. As is mentioned above,
these results are true if the masses of produced particles are small
in comparison with $m_\Phi$. One should keep in mind that the former includes
the time dependent contribution from the interaction with 
$\Phi$ (\ref{l-int}):
\be
m_\psi^{(eff)} = m_0 + g \Phi_0 (t) \cos \left( m_\Phi t + \delta \right),
\label{m-eff}
\ee
Due to the last term the effective mass of the produced
particles rises together with rising amplitude of the classical 
$\Phi$. Because of that the probability of particle production by this
field drops down for large field amplitude, a counter-intuitive result. 
If $g\Phi_0 > m_\Phi$, the perturbative
approach  becomes inapplicable and non-perturbative calculations are
necessary. In this case, and for $m_0<m_\Phi$, the productions of fermions 
is suppressed as~\cite{ad-dk}: 
\be
\Gamma \sim \Gamma_{pt} \left( \frac{m_\Phi}{g\Phi_0}\right)^{1/2}
\label{gam-npt}
\ee
Production takes place during a short time interval when the effective 
mass of $\psi$ is close to zero. This gives rise to a power law 
suppression. However, if $m_0 \gg m_\Phi$ then the particle production 
is exponentially suppressed,
$\sim \exp[-m_0/m_{\Phi}]$. If $m_0>m_\Phi$, but not too much bigger, then
the suppression of production would be suppressed by an additional power in 
the coupling constant because the multi-quantum production would be 
necessary to respect the energy conservation law. 

However, the production of would-be-heavy particles is quite efficient 
if the same trick as described above is realized: namely, the bare masses of
particles are very small initially but after they are produced they acquire
large masses by a Higgs type phase transition. 
On the other hand, as is argued e.g. in ref.~\cite{heavy-f} heavy fermions
with masses up to $10^{17}-10^{18}$ GeV could be efficiently produced
at the preheating stage, even without such a trick. This statement contradicts
the results of ref.~\cite{ad-dk}, according to which heavy fermions are 
always much slower produced than massless ones. 

Production of bosons could be strongly amplified by the parametric 
resonance~\cite{ad-dk,tb,kls}. In quantum language
this resonance can be understood
as Bose condensation of the produced particles in the mode
$p^2 = m^2_\Phi/4 - m^2_\chi$ (for the coupling $\Phi \chi \chi^*$).  
The resonance should be sufficiently wide to keep the produced particles in 
the resonance mode despite red-shift and possible (self)scattering.

\subsection{Gravitational particle production \label{ss-grav-prod}}

As we mentioned above, a time varying external field can produce elementary 
particles. This should be true in particular for gravitational field. 
However, the FRW metric is known to be conformally flat, i.e.
after suitable redefinition of coordinates the interval can be
rewritten in Minkowskian form with a common scale factor:
\be
ds^2 = a^2(\eta,{\bf r}) \left( d\eta^2 - d{\bf r}^2 \right)
\label{ds-conf}
\ee
where in spatially flat universe the scale factor $a$ depends only on
conformal time $\eta$ and not on the space coordinate ${\bf r}$.

From this property it immediately  follows that conformally 
invariant particles, e.g. massless fermions or vector bosons are not
created by cosmological gravitational field~\cite{parker}. This is true
at least classically but quantum corrections, leading to well known
triangle trace anomaly\cite{ch-el}, 
break conformal invariance and allows for
production of massless gauge bosons in FRW background~\cite{ad-anom}.
Of course massive particles are not conformally invariant and their
production rate is proportional to $m_x^2$. Calculations of massive
particle production in cosmological background have been done in the 
70th~\cite{grav-prod} and are reviewed in the book~\cite{grib-book}.
Recently, because of the renewed interest to heavy particle production 
in cosmology, more publications appeared~\cite{new-grav}  and has been
reviewed~\cite{kuz-tka}. 

According to the calculations of the quoted papers, the ratio of the 
energy densities of the produced particles to the total cosmological
energy density is 
\be
\rho_x /\rho_{tot} \sim (\mx/\mpl)^2
\label{rhox}
\ee
For comparison particle production due to the trace anomaly gives
\be
\rho_{anom}/\rho_{tot} \sim \left( \alpha \beta \right)^2 
\left (H_{prod} / \mpl \right)^2
\label{rho-anom}
\ee
where $\alpha \sim 1/50$ is the fine structure constant at high energies,
$\beta$ is the leading coefficient of the beta-function, and $H_{prod}$ is
the value of the Hubble constant at the earliest moment of the production;
in inflationary model it should be the Hubble parameter at the end of 
inflation. Anomalous production is operative only for light particles
with masses smaller than $H_{prod}$. If these particles acquire a large
mass later, after some phase transition, anomalous production could be 
a dominant cosmological source of heavy particles.

Theoretical description of the particle production in cosmology encounters 
the following problem. The notion of particle depends upon the definition of 
the vacuum state and particle production  is unambiguously described if
external field which produces particles disappears at positive and negative
time infinities. In this case one can define positive and negative energy
states corresponding to particles and antiparticles and use the standard
technique with the Bogolyubov coefficients. However, in cosmology the
gravitational field does not disappear in initial state, even more it
is typically very strong, and definition of vacuum meets serious problems.
These problems are related to the nonlocal character of separating of
positive and negative energies. In particular, the particle number
operator is a nonlocal one.
To avoid these problem one may work with local operators e.g. with the
energy-momentum tensor, $T_{\mu\nu}$.
The definitions of pressure and energy densities are unambiguous and 
one can calculate the time evolution of the latter and a modification of
the equation of state of cosmological matter which in turn leads to
a change in the expansion regime.

It is interesting in this connection to mention the Unruh
effect~\cite{unruh}: an accelerated observer in vacuum would see
locally thermal bath of particles with temperature proportional to
acceleration. On the other hand if we calculate the total 
energy-momentum tensor of the system it should be identically zero. It 
is evidently zero in the original inertial frame and it remains
zero after we make a transformation to the accelerated frame. In these
terms the effect is simply redistribution of the total local $T_{\mu\nu}$
between the vacuum, $T_{\mu\nu}^{(vac)}$, and particle,  
$T_{\mu\nu}^{(part)}$, ones in the accelerated frame such that 
$T_{\mu\nu}^{(vac)}+ T_{\mu\nu}^{(part)}=0$.

\section{Cosmological impact of HqSP \label{cosm-eff}}

If the life time of heavy particles is smaller than 1 sec, then they
would not play any noticeable role in big bang nucleosynthesis (BBN)
and would be practically unnoticeable. The decays of HqSP could create
baryon asymmetry of the universe but it is impossible to verify. For
larger life-times the effects of their decays could be observable in
BBN, CMBR, large scale structure, and, as is mentioned above, in 
ultra high energy cosmic rays. 
With very long life-time, $\taux > T_U$, 
HqSP could be dark matter particles. Their number density in this case
is bounded from above by 
\be
\mx n_x < 1.5 \,{\rm keV/cm}^3
\label{mx-nx}
\ee
i.e. $n_x < 10^{-20}/{\rm cm}^3 (10^{14}\,{\rm GeV}/\mx) $ or
equal to that if these particles dominate cosmological dark matter.

The frozen number density of heavy particles can be estimated as
\be
n_x/n_\gamma \sim \left[V_x \sigma_x^{(ann)} \mx \mpl \right]^{-1},
\label{nx-ngam}
\ee
where $n_\gamma = 410/{\rm cm}^3$ is the number density of CMBR photons, 
$\sigma_x$ is the cross-section of $x\bar x$-annihilation and $V_x$ is the
c.m. velocity of $x$-particles. To satisfy the constraint 
(\ref{mx-nx}) either the cross-section should be very large:
\be
\sigma_x V_x > 10^{-38} {\rm cm}^2 = 
10^{18} \, \left( \mx /10^{14}\,{\rm GeV} \right)
\label{sigma-x}
\ee
or the production of $x$ in the early universe must be strongly suppressed,
so the initial number density was much below the equilibrium value. 

Interesting bounds on the properties of the super-heavy dark matter
particles, come from consideration of the character of the density 
perturbations~\cite{dkrs}. For non-equilibrium super-heavy dark matter
isocurvature perturbations are non-negligible and since the latter are
bounded by CMBR to be at most at the level of 10\% with respect to the 
total density perturbations, this allows to conclude that $\mx$
must be larger than the Hubble parameter at the end of inflation. Some 
model dependent bounds on the reheating temperature has been also
derived~\cite{dkrs}.

\section{Conclusion \label{s-concl}}

There is no compelling reason to expect that 
heavy quasi-stable particles exist.
Of course grand unified theories (GUT) predict super-heavy gauge bosons 
with $m_{GUT}\sim 10^{15}-10^{16}$ GeV but they have very short 
life-time, $\tau \sim (\alpha m_{GUT} )^{-1}$. 
Magnetic monopoles might exist but they should be absolutely stable.
Stability of the Planck mass black holes is questionable. Thus, if heavy
elementary particles exist they are most probably unstable. 
Stability could be ensured by a new symmetry and some conserved 
quantum numbers. However, a global symmetry, if broken by Planck scale 
physics, would not be able to make life-time sufficiently long. A new 
local symmetry, e.g. new QCD near the Planck scale could be broken
very weakly, in contrast to global symmetries, and the life-time 
of $x$ might be long.

With TeV gravity (quasi)stability is even more difficult to achieve
but a new local symmetry with slightly massive gauge bosons could
naturally allow for a large life-time.

It is difficult to confirm or reject the hypothesis of cosmological
HqSP. Ultrahigh energy cosmic rays from their decay or annihilation
could present a positive evidence if other possible explanations are 
excluded. It is even more difficult to establish if HqSP or 
some other objects make the bulk 
of cosmological dark matter. At the present day there are no definitive
features in large scale structure or CMBR that could be reliably identified
as signatures of heavy stable particles. Still though the chances are small
the stakes are high. An importance of astronomical discovery of very heavy 
particles, unaccessible to existing accelerators even in foreseeable future, 
is difficult to overestimate.

\end{document}